%% file: bbhpaper.tex
\pdfoutput=1
\documentclass[prd,
               twocolumn,
               amssymb,
               amsmath,
               eqsecnum,
               showpacs,
               letterpaper,
               superscriptaddress,
               altaffilletter,
               floatfix]
               {revtex4}
\usepackage{graphicx}
\usepackage{amssymb}
\usepackage{multirow}
\usepackage{color}

\newcommand{\e}[1]{$\times10^{#1}$}

\newcommand{\msun}{M${}_{\odot}\,$}
\newcommand{\Gcyr}{GC${}^{-1}$Gyr${}^{-1}\,$}

\newcommand{\MpcMyr}{Mpc${}^{-3}$Myr${}^{-1}\,$}

\newcommand{\bfxi}{\boldsymbol\xi}

\newcommand{\VV}{ V_{\textrm{vis}} }
\newcommand{\VVavg}{ \bar{V}_{\textrm{vis}} }

\begin{document}

\title{Search for Gravitational Waves from Intermediate Mass Binary Black Holes}

\begin{abstract}
We present the results of a weakly modeled burst search for gravitational waves from mergers of non-spinning intermediate mass black holes (IMBH) in the total mass range 100--450~\msun and with the component mass ratios between 1:1 and 4:1. The search was conducted on data collected by the LIGO and Virgo detectors between November of 2005 and October of 2007. No plausible signals were observed by the search which constrains the astrophysical rates of the IMBH mergers as a function of the component masses. In the most efficiently detected bin centered on $88+88$~\msun, for non-spinning sources, the rate density upper limit is 0.13 per Mpc${}^3$ per Myr at the $90$\% confidence level.
\end{abstract}



\affiliation{LIGO - California Institute of Technology, Pasadena, CA  91125, USA$^\ast$}
\affiliation{California State University Fullerton, Fullerton CA 92831 USA$^\ast$}
\affiliation{SUPA, University of Glasgow, Glasgow, G12 8QQ, United Kingdom$^\ast$}
\affiliation{Laboratoire d'Annecy-le-Vieux de Physique des Particules (LAPP), Universit\'e de Savoie, CNRS/IN2P3, F-74941 Annecy-Le-Vieux, France$^\dagger$}
\affiliation{INFN, Sezione di Napoli $^a$; Universit\`a di Napoli 'Federico II'$^b$ Complesso Universitario di Monte S.Angelo, I-80126 Napoli; Universit\`a di Salerno, Fisciano, I-84084 Salerno$^c$, Italy$^\dagger$}
\affiliation{LIGO - Livingston Observatory, Livingston, LA  70754, USA$^\ast$}
\affiliation{Albert-Einstein-Institut, Max-Planck-Institut f\"ur Gravitationsphysik, D-30167 Hannover, Germany$^\ast$}
\affiliation{Leibniz Universit\"at Hannover, D-30167 Hannover, Germany$^\ast$}
\affiliation{Nikhef, Science Park, Amsterdam, the Netherlands$^a$; VU University Amsterdam, De Boelelaan 1081, 1081 HV Amsterdam, the Netherlands$^b$$^\dagger$}
\affiliation{National Astronomical Observatory of Japan, Tokyo  181-8588, Japan$^\ast$}
\affiliation{University of Wisconsin--Milwaukee, Milwaukee, WI  53201, USA$^\ast$}
\affiliation{University of Florida, Gainesville, FL  32611, USA$^\ast$}
\affiliation{University of Birmingham, Birmingham, B15 2TT, United Kingdom$^\ast$}
\affiliation{INFN, Sezione di Roma$^a$; Universit\`a 'La Sapienza'$^b$, I-00185 Roma, Italy$^\dagger$}
\affiliation{LIGO - Hanford Observatory, Richland, WA  99352, USA$^\ast$}
\affiliation{Albert-Einstein-Institut, Max-Planck-Institut f\"ur Gravitationsphysik, D-14476 Golm, Germany$^\ast$}
\affiliation{Montana State University, Bozeman, MT 59717, USA$^\ast$}
\affiliation{European Gravitational Observatory (EGO), I-56021 Cascina (PI), Italy$^\dagger$}
\affiliation{Syracuse University, Syracuse, NY  13244, USA$^\ast$}
\affiliation{LIGO - Massachusetts Institute of Technology, Cambridge, MA 02139, USA$^\ast$}
\affiliation{Laboratoire AstroParticule et Cosmologie (APC) Universit\'e Paris Diderot, CNRS: IN2P3, CEA: DSM/IRFU, Observatoire de Paris, 10 rue A.Domon et L.Duquet, 75013 Paris - France$^\dagger$}
\affiliation{Columbia University, New York, NY  10027, USA$^\ast$}
\affiliation{INFN, Sezione di Pisa$^a$; Universit\`a di Pisa$^b$; I-56127 Pisa; Universit\`a di Siena, I-53100 Siena$^c$, Italy$^\dagger$}
\affiliation{Stanford University, Stanford, CA  94305, USA$^\ast$}
\affiliation{IM-PAN 00-956 Warsaw$^a$; Astronomical Observatory Warsaw University 00-478 Warsaw$^b$; CAMK-PAN 00-716 Warsaw$^c$; Bia{\l}ystok University 15-424 Bia{\l}ystok$^d$; IPJ 05-400 \'Swierk-Otwock$^e$; Institute of Astronomy 65-265 Zielona G\'ora$^f$,  Poland$^\dagger$}
\affiliation{The University of Texas at Brownsville and Texas Southmost College, Brownsville, TX  78520, USA$^\ast$}
\affiliation{San Jose State University, San Jose, CA 95192, USA$^\ast$}
\affiliation{Moscow State University, Moscow, 119992, Russia$^\ast$}
\affiliation{LAL, Universit\'e Paris-Sud, IN2P3/CNRS, F-91898 Orsay$^a$; ESPCI, CNRS,  F-75005 Paris$^b$, France$^\dagger$}
\affiliation{NASA/Goddard Space Flight Center, Greenbelt, MD  20771, USA$^\ast$}
\affiliation{University of Western Australia, Crawley, WA 6009, Australia$^\ast$}
\affiliation{The Pennsylvania State University, University Park, PA  16802, USA$^\ast$}
\affiliation{Universit\'e Nice-Sophia-Antipolis, CNRS, Observatoire de la C\^ote d'Azur, F-06304 Nice$^a$; Institut de Physique de Rennes, CNRS, Universit\'e de Rennes 1, 35042 Rennes$^b$, France$^\dagger$}
\affiliation{Laboratoire des Mat\'eriaux Avanc\'es (LMA), IN2P3/CNRS, F-69622 Villeurbanne, Lyon, France$^\dagger$}
\affiliation{Washington State University, Pullman, WA 99164, USA$^\ast$}
\affiliation{INFN, Sezione di Perugia$^a$; Universit\`a di Perugia$^b$, I-06123 Perugia,Italy$^\dagger$}
\affiliation{INFN, Sezione di Firenze, I-50019 Sesto Fiorentino$^a$; Universit\`a degli Studi di Urbino 'Carlo Bo', I-61029 Urbino$^b$, Italy$^\dagger$}
\affiliation{University of Oregon, Eugene, OR  97403, USA$^\ast$}
\affiliation{Laboratoire Kastler Brossel, ENS, CNRS, UPMC, Universit\'e Pierre et Marie Curie, 4 Place Jussieu, F-75005 Paris, France$^\dagger$}
\affiliation{University of Maryland, College Park, MD 20742 USA$^\ast$}
\affiliation{University of Massachusetts - Amherst, Amherst, MA 01003, USA$^\ast$}
\affiliation{Canadian Institute for Theoretical Astrophysics, University of Toronto, Toronto, Ontario, M5S 3H8, Canada$^\ast$}
\affiliation{Tsinghua University, Beijing 100084 China$^\ast$}
\affiliation{University of Michigan, Ann Arbor, MI  48109, USA$^\ast$}
\affiliation{Louisiana State University, Baton Rouge, LA  70803, USA$^\ast$}
\affiliation{The University of Mississippi, University, MS 38677, USA$^\ast$}
\affiliation{Charles Sturt University, Wagga Wagga, NSW 2678, Australia$^\ast$}
\affiliation{Caltech-CaRT, Pasadena, CA  91125, USA$^\ast$}
\affiliation{INFN, Sezione di Genova;  I-16146  Genova, Italy$^\dagger$}
\affiliation{Pusan National University, Busan 609-735, Korea$^\ast$}
\affiliation{Australian National University, Canberra, ACT 0200, Australia$^\ast$}
\affiliation{Carleton College, Northfield, MN  55057, USA$^\ast$}
\affiliation{The University of Melbourne, Parkville, VIC 3010, Australia$^\ast$}
\affiliation{Cardiff University, Cardiff, CF24 3AA, United Kingdom$^\ast$}
\affiliation{INFN, Sezione di Roma Tor Vergata$^a$; Universit\`a di Roma Tor Vergata, I-00133 Roma$^b$; Universit\`a dell'Aquila, I-67100 L'Aquila$^c$, Italy$^\dagger$}
\affiliation{University of Salerno, I-84084 Fisciano (Salerno), Italy and INFN (Sezione di Napoli), Italy$^\ast$}
\affiliation{The University of Sheffield, Sheffield S10 2TN, United Kingdom$^\ast$}
\affiliation{RMKI, H-1121 Budapest, Konkoly Thege Mikl\'os \'ut 29-33, Hungary$^\dagger$}
\affiliation{INFN, Gruppo Collegato di Trento$^a$ and Universit\`a di Trento$^b$,  I-38050 Povo, Trento, Italy;   INFN, Sezione di Padova$^c$ and Universit\`a di Padova$^d$, I-35131 Padova, Italy$^\dagger$}
\affiliation{Inter-University Centre for Astronomy and Astrophysics, Pune - 411007, India$^\ast$}
\affiliation{California Institute of Technology, Pasadena, CA  91125, USA$^\ast$}
\affiliation{Northwestern University, Evanston, IL  60208, USA$^\ast$}
\affiliation{University of Cambridge, Cambridge, CB2 1TN, United Kingdom$^\ast$}
\affiliation{The University of Texas at Austin, Austin, TX 78712, USA$^\ast$}
\affiliation{E\"otv\"os Lor\'and University, Budapest, 1117 Hungary$^\ast$}
\affiliation{University of Szeged, 6720 Szeged, D\'om t\'er 9, Hungary$^\ast$}
\affiliation{Universitat de les Illes Balears, E-07122 Palma de Mallorca, Spain$^\ast$}
\affiliation{Rutherford Appleton Laboratory, HSIC, Chilton, Didcot, Oxon OX11 0QX United Kingdom$^\ast$}
\affiliation{Embry-Riddle Aeronautical University, Prescott, AZ   86301 USA$^\ast$}
\affiliation{National Institute for Mathematical Sciences, Daejeon 305-390, Korea$^\ast$}
\affiliation{Perimeter Institute for Theoretical Physics, Ontario, N2L 2Y5, Canada$^\ast$}
\affiliation{University of New Hampshire, Durham, NH 03824, USA$^\ast$}
\affiliation{University of Adelaide, Adelaide, SA 5005, Australia$^\ast$}
\affiliation{University of Southampton, Southampton, SO17 1BJ, United Kingdom$^\ast$}
\affiliation{University of Minnesota, Minneapolis, MN 55455, USA$^\ast$}
\affiliation{Korea Institute of Science and Technology Information, Daejeon 305-806, Korea$^\ast$}
\affiliation{Hobart and William Smith Colleges, Geneva, NY  14456, USA$^\ast$}
\affiliation{Institute of Applied Physics, Nizhny Novgorod, 603950, Russia$^\ast$}
\affiliation{Lund Observatory, Box 43, SE-221 00, Lund, Sweden$^\ast$}
\affiliation{Hanyang University, Seoul 133-791, Korea$^\ast$}
\affiliation{Seoul National University, Seoul 151-742, Korea$^\ast$}
\affiliation{University of Strathclyde, Glasgow, G1 1XQ, United Kingdom$^\ast$}
\affiliation{Southern University and A\&M College, Baton Rouge, LA  70813, USA$^\ast$}
\affiliation{University of Rochester, Rochester, NY  14627, USA$^\ast$}
\affiliation{Rochester Institute of Technology, Rochester, NY  14623, USA$^\ast$}
\affiliation{University of Sannio at Benevento, I-82100 Benevento, Italy and INFN (Sezione di Napoli), Italy$^\ast$}
\affiliation{Louisiana Tech University, Ruston, LA  71272, USA$^\ast$}
\affiliation{McNeese State University, Lake Charles, LA 70609 USA$^\ast$}
\affiliation{Andrews University, Berrien Springs, MI 49104 USA$^\ast$}
\affiliation{Trinity University, San Antonio, TX  78212, USA$^\ast$}
\affiliation{Southeastern Louisiana University, Hammond, LA  70402, USA$^\ast$}

\author{J.~Abadie$^\text{1}$}\noaffiliation\author{B.~P.~Abbott$^\text{1}$}\noaffiliation\author{R.~Abbott$^\text{1}$}\noaffiliation\author{T.~D.~Abbott$^\text{2}$}\noaffiliation\author{M.~Abernathy$^\text{3}$}\noaffiliation\author{T.~Accadia$^\text{4}$}\noaffiliation\author{F.~Acernese$^\text{5a,5c}$}\noaffiliation\author{C.~Adams$^\text{6}$}\noaffiliation\author{R.~Adhikari$^\text{1}$}\noaffiliation\author{C.~Affeldt$^\text{7,8}$}\noaffiliation
% ...

\collaboration{$^\ast$The LIGO Scientific Collaboration and $^\dagger$The Virgo Collaboration}
\noaffiliation


\date[\relax]{Dated: \today}

\maketitle

\input{intro}

\input{experiment}

\input{simulations}

\input{results}

\input{summary}

\input{acknowledgements}

\appendix
\input{appendix_ner}

\bibliographystyle{apsrev}
\bibliography{bbhpaper}

\end{document}

%% file: intro.tex
\section{Introduction}

Emission of gravitational waves (GW) via strong general relativistic processes between two compact objects (black holes and/or neutron stars) is the hallmark of compact binary coalescence (CBC). Binary black holes, a particular class of CBC sources, have been one of the main detection targets of ground based gravitational wave detectors since the inception of large wide-band interferometers~\cite{0034-4885-72-7-076901,0264-9381-23-19-S01,Grote:2008}. This paper presents the results of a search for gravitational waves from the coalescence of intermediate mass black holes (IMBH). The search used data collected by the Laser Interferometer Gravitational-Wave Observatory (LIGO) during its fifth science run (S5) from November 2005 to October 2007~\cite{0034-4885-72-7-076901} and by the Virgo GW interferometer~\cite{Acernese:2008kx}, which commenced its first science run (VSR1) in May 2007 and operated in coincidence with LIGO. 

The coalescence of compact binaries is generally divided into three stages: the inspiral, merger, and ringdown. Gravitational waves from the inspiral stage are quasi-periodic ``chirp'' signals of increasing frequency and amplitude which are well described by analytical post-Newtonian (PN) models~\cite{Blanchet:2001og,Blanchet:2002av,Boyle:2007ft, PhysRevD.51.5360} before the binary evolution reaches the inner-most stable circular orbit (ISCO). Near the ISCO, the strong gravitational interaction no longer allows for a stable orbit and the two black holes merge together to form a single black hole. After the merger stage, the newly born perturbed black hole emits gravitational waves via exponentially damped quasi-normal modes in the ringdown stage. The merger and ringdown stages of the GW signal are important for detection of IMBH sources because the characteristic frequencies of the inspiral stage are usually outside of the sensitivity band of ground-based GW interferometers.  Recent progress in numerical relativity (NR) has expanded the understanding of binary black hole systems through the merger and ringdown stages~\cite{Buonanno:1999sp,Damour:2008lu,Ajith:2007fj,Buonanno:2007sf,Buonanno:2009sf,Buonanno:2007qm} allowing calculation of the full inspiral-merger-ringdown (IMR) waveforms. 

Several matched-filter searches have been conducted for CBC sources consisting of total masses less than 35~\msun~\cite{Abbot:2007uq,Abbott:2009bh,Collaboration:2009fy} with inspiral templates. One more search in the total mass range 25-100~\msun (where the contribution of the inspiral stage is still dominant) was performed with IMR templates~\cite{Collaboration:2011fk}. In order to identify GW events in the noisy data, these searches rely on the generation of template banks from the signal model. Currently, the generation of complete and accurate template banks in the mass region above 100~\msun is challenging. Therefore, for the IMBH search reported here we used the Coherent WaveBurst algorithm~\cite{Abbott:2008eh,LIGO:2009pz,Abadie:2010xk} which is designed for detection of un-modeled burst signals and does not require \emph{a priori} knowledge of the signal waveforms. However, due to the lack of model constraints, generic burst searches are usually more affected by the background than matched-filter searches. To improve the rejection of background events, the CWB algorithm can enforce a constraint on the waveform polarization~\cite{Pankow:2009lv}. Such a constrained burst algorithm can be used to search for IMBH coalescences without the need of template banks while still achieving nearly the same detection sensitivity.

\subsection{Intermediate Mass Black Hole Formation}

IMBHs have been posited to complete the black hole mass hierarchy. As such, IMBHs cover several decades in the black hole mass spectrum between stellar mass black holes of a few tens of~\msun, formed from star collapse, and super-massive black holes of $10^5$ \msun or more present in the center of galaxies. Some models of IMBH formation include runaway stellar collision scenarios~\cite{0004-637X-576-2-899} in globular clusters (GC). One model proposes that lower mass single IMBHs could be formed by the stalled supernova of early Population III stars~\cite{2004IJMPD..13....1M,0004-637X-706-2-1184}. Another model~\cite{Miller:2002uq} studies the progressive accumulation of mass into a large ($>50$ \msun) seed black hole via coalescence of a population of smaller black holes. However, the existence of binaries with IMBH components remains uncertain since stellar winds may stall the growth of the IMBH progenitors in the runaway collision scenario~\cite{refId}, or the merger recoil may also eject a newly formed black hole 
from the globular cluster~\cite{0004-637X-613-2-1143,0004-637X-637-2-937}.

IMBHs have been searched for via conventional astronomy, and a few candidates exist~\cite{Strohmayer:2009of}. It has been suggested that IMBHs are the engines powering ultraluminous X-ray (ULX) sources~\cite{Colbert:2004sz,Zampieri:2010uq} such as M82 X-1~\cite{1538-4357-652-2-L105,Casella:2008fk} or NGC 1313 X-2~\cite{Patruno:2010uq}. Most models agree that the primary hosts of these objects would be globular clusters~\cite{0004-637X-620-1-238,Safonova:2009qp,Vesperini:2010kx}. These objects are thought to grow from accretion of smaller compact objects~\cite{0004-637X-616-1-221}, and therefore, IMBHs could be a prime candidate for the detection of GW by the coalescence of solar mass objects into the central BH. The detection of an IMBH binary would not just represent the first detection of GW, but could have important consequences for theories about the formation of super-massive black holes and the dynamics and evolution of globular clusters~\cite{Umbreit:2009nt,Vesperini:2010kx}. 

In this search, we focus on the IMBH binary systems with total masses between 100-450 \msun and component black hole masses with the mass ratios between 1:1 and 4:1. The expected GW emission from these sources is in the frequency band between tens and few hundred Hz. This frequency band includes the most sensitive band of the initial ground based GW detectors. Those IMBH systems considered in this search contain most of the detectable signal power in the merger and ringdown because the power emitted during the earlier inspiral stage is in the frequency band below 40~Hz where there is a rapid deterioration of the LIGO/Virgo network sensitivity due to seismic noise. Above 450 \msun, the power emitted is no longer present at accessible frequencies. 

Concerning the rate of IMBH--IMBH coalescence, the upper limit has been estimated at 0.07 \Gcyr~\cite{ratesdoc}. Using an astrophysical source density of 0.3 GC Mpc${}^{-3}$~\cite{Mandel:2007rates}, this corresponds to 2\e{-5}~\MpcMyr. If intermediate mass ratio (mass ratios of 10:1 or greater) inspirals onto IMBH are also considered, the rate is conceivably as high as 3~\Gcyr (9\e{-4}~\MpcMyr). The \emph{detection} rate estimates for the IMBH systems considered in this search are much smaller than 1 yr${}^{-1}$. However, predicted detection rates for second generation detectors such as Advanced LIGO and Virgo increase by orders of magnitude over initial detectors as their proposed designs include better sensitivity at comparatively lower frequencies.

%% file: experiment.tex
\section{Experiment}
\label{sec:experiment}

Five GW detectors were operating during the S5/VSR1 runs: two detectors (4~km detector H1 and 2~km detector H2) at the LIGO site in Hanford, Washington, another 4~km LIGO detector (L1) in Livingston, Louisiana, the 3~km Virgo detector (V1) in Cascina, Italy, and the 600~m GEO600 detector in Hannover, Germany. The GEO600 detector had a significantly lower sensitivity to the IMBH sources than the other four detectors and therefore it was not considered in this search. Due to limited detector duty cycles, there were several network configurations consisting of two to four detectors operating in coincidence. In this search we considered two networks with the most accumulated observation time: the three-fold network L1H1H2 and the four-fold network L1H1H2V1. 

Not all data which was collected by the detectors is used in the analysis. Extensive studies~\cite{Blackburn:2008rt,VirgoDetChar} have been performed to identify (flag) data segments with high seismic activity, large mechanical disturbances, and a high rate of environmental and instrumental transients. These data quality flags are nearly identical to those used for the S5/VSR1 all-sky burst analysis~\cite{LIGO:2009pz,Abadie:2010xk}. Data quality flags are classified into different categories, starting with the initial flags selecting data segments used by the search algorithm. Further data quality flags are imposed on all events emerging from the search algorithm, including a set of event vetoes derived from well known correlations between the GW data channel and the auxiliary channels. All events passing these checks are considered as detection candidates. Finally, a set of data quality flags is used to remove events with weaker environmental and instrumental correlations. The set of events passing the final checks is then used for estimation of the astrophysical rate limits. Table~\ref{tbl:livetime} shows the total observation time for the network configurations used in the search after all data quality flags are applied.

\begin{table}[htbp]
\begin{center}
\caption{Summary of each network's analyzed observation time after all data quality flags are applied.}
\begin{tabular}{lc}
\hline
detector network & observation time (yr)  \\ 
\hline \hline
L1H1H2V1 & 0.16 \\ 
L1H1H2 & 0.65 \\ 
\hline
\end{tabular}
\label{tbl:livetime}
\end{center}
\end{table}

\section{Intermediate Mass Binary Black Hole Search}

\subsection{Search Algorithm}
\label{sec:dapipe}

The IMBH search is based on the Coherent WaveBurst (CWB) algorithm~\cite{Klimenko:2007hd} which has been used in the S5/VSR1 burst searches~\cite{LIGO:2009pz,Abadie:2010xk}. The CWB algorithm performs a constrained likelihood analysis~\cite{Klimenko:2005wa} of the network data stream, reconstructing detector responses to an anticipated GW signal. The residual data (null stream), obtained after subtraction of the estimated detector responses from the original data, represents the reconstructed network noise. Along with the reconstruction of un-modeled burst signals, which imply random polarization, CWB performs likelihood analysis of signals with other polarization states, including elliptical, linear, and circular polarizations. 

In this search, we use the elliptical polarization constraint.  The details of the likelihood analysis and the elliptical polarization constraint are presented in Appendix~\ref{elliptical}. Though not completely generic, the constraint improves rejection of those background events originating from the random coincidence of the environmental and instrumental transients in the detectors. In general, the signal polarization may evolve as a function of time. For example, spinning black hole systems have slowly evolving, large polarization changes that track the precession of the orbital angular momentum~\cite{Apostolatos:1994}. Even non-spinning black hole systems have small, rapidly oscillating polarization changes as different multipolar orders interfere constructively and destructively~\cite{Blanchet:1996pi}. Either case will introduce a time dependence on the waveform polarization. However, the effects of the spin-orbit 
coupling become significant only for a subset of black hole systems with spinning components. Even for these signals, a significant fraction of the band-limited power can be associated with some instantaneous polarization. Thus, the constraint should not significantly affect the detection efficiency of IMBH sources.

Three major statistics --- obtained as the result of the likelihood analysis --- are used for selection of reconstructed events: the network correlation coefficient $cc$, the network energy disbalance $\lambda_{\mathrm{net}}$ and the coherent network amplitude $\eta$ (see Appendix~\ref{elliptical}). The statistics $cc$ and $\lambda_{\mathrm{net}}$ are used to characterize the conformance of identified events with the signal model and its constraints. A low value of $cc\ll 1$ is typical for background events which tend to have a large residual energy and a small coherent energy of reconstructed signals. On the contrary, a genuine GW event is characterized by a value of $cc$ close to unity. The energy disbalance $\lambda_{\mathrm{net}}$ identifies the un-physical solutions of the likelihood functional which are typical for spurious events. A significant deviation of $\lambda_{\mathrm{net}}$ from zero is an indication that the energy of the reconstructed response is significantly larger than the energy of the data stream in at least one detector. Table~\ref{tbl:cuts} shows the $cc$ and $\lambda_{\mathrm{net}}$ thresholds used in the analysis.

\begin{table}[htbp]
\centering
\caption{Post-production selection cuts: candidate events are selected if
$\lambda_{\mathrm{net}}$ and $cc$ are, respectively, less and greater than indicated thresholds.}
\begin{tabular}{lcc}
\hline 
network & H1H2L1V1 & H1H2L1 \\
\hline \hline
dual stream energy disbalance ($\lambda_{\mathrm{net}}$) & 0.2 & 0.15 \\
network correlation coefficient (cc) & 0.6 & 0.70 \\
\hline
\end{tabular}
\label{tbl:cuts}
\end{table}

The coherent network amplitude $\eta$ is the main CWB detection statistic. It is proportional to the signal-to-noise ratio (SNR) and is used to rank selected events and establish their significance against a sample of background events. 

\subsection{Background Estimation}

We estimate the false alarm rate of events originating from the detector noise by introducing artificial time shifts (far exceeding the intersite light travel time) between the data from different sites before using the search algorithm. This procedure assumes that the noise induced events are not correlated between the sites. Events obtained from the time-shifted data represent the search background sample. Data from different detectors is shifted by integer multiples of one second per time-shift configuration (time lag). The H1H2L1 network had a total of 600 time lags performed, and the H1H2L1V1 network had a total of 1000 time lags including the foreground (zero lag) configuration. In total, this procedure accumulated 569 years of effective background live time for the three detector network and 180 years of background live time for the four detector network. The background events that survived the data quality and the analysis selection cuts (see Table~\ref{tbl:cuts}) are used for calculation of the significance of candidate events and  the false alarm density statistic described in section~\ref{sec:fad}.

%% file: simulations.tex
\section{Simulations}
\label{sec:simulation}

To characterize the detection efficiency of the search in the parameter space of potential IMBH sources, extensive simulation studies were performed with different families of the IMR waveforms. These studies were made to determine a sensitivity volume of the search, also called visible volume, assuming that the IMBH sources are distributed uniformly in space. In order to calculate the visible volume a Monte Carlo detection efficiency study was performed by adding into the data waveforms drawn randomly from the physical parameter space which we consider. The simulated detector responses were injected via software into detector data and the search algorithm was used to identify the injections. A large sample of waveforms for each network configuration was generated to sufficiently cover the parameter space of the IMBH sources presented in Table~\ref{tbl:params}. The simulated waveforms were distributed in a spherical volume with a radius of 2 Gpc and a uniform distribution over the source inclination and polarization angles, and sky locations.

\begin{table}[htbp]
\begin{center}
\caption{Summary of injected waveform parameters.}
\begin{tabular}{lr}
\hline
Total Mass (\msun) & 100 -- 450  \\
Mass Ratio & 1 -- 4  \\ 
Distance (Mpc) & 0 -- 2000 \\
\hline
\end{tabular}
\label{tbl:params}
\end{center}
\end{table}

In this simulation, redshift corrections were neglected because very few injections are placed (and detected) at the distances which would require consideration of this effect. Spin of the component black holes was not considered as well, but a discussion of potential spin effects is presented in section~\ref{sec:sum}. 
\subsection{Simulated Waveforms}

Most of the previous template based searches~\cite{LIGO:2009pz,Collaboration:2009fy,Abbott:2009bh,LSC:2010fk} used only inspiral or ringdown templates~\cite{Goggin:2009fv} to do simulation studies. As the total mass of the system increases, the analytical PN inspiral waveforms become inadequate because only the merger and ringdown waves have significant power in the sensitive band of the detectors. For this reason, this search (and a template search for binary black holes with the total mass between 25 and 100 \msun~\cite{Collaboration:2011fk}) uses the full IMR waveforms from two different families: the Effective One Body Numerical Relativity (EOBNR) family~\cite{Buonanno:1999sp,Buonanno:2007sf,Damour:2008lu,Buonanno:2009qa} and the IMRPhenom family~\cite{Ajith:2007fj,Ajith:2007qp,Ajith:2009gj}. The EOBNR waveform family uses the Effective One Body (EOB) Hamiltonian to evolve the binary system up to the merger. The EOB approach is able to simulate the dynamics of the plunge into merger of the binary black hole system through 3 PN order. Further accuracy has been obtained by the use of ``pseudo'' 4 PN terms motivated from the results of numerical relativity simulations. To complete the evolution from the plunge-merger to ringdown, a superposition of the ringdown frequency modes is matched to the end of the merger. The IMRPhenom family is constructed by matching 3.5 PN order analytical inspiral waveforms to the corresponding NR merger waveforms  to make ``hybrid'' waveforms. These hybrid waveforms are then extrapolated to form a full waveform family in the Fourier domain. In contrast to \cite{Collaboration:2011fk} which constructed template banks from the EOBNR family but used both families for detection efficiency studies, this search uses only the EOBNR family for efficiency studies. While not used for a detailed simulation, the IMRPhenom family was used in the IMBH analysis to cross-check the validity of results obtained with the EOBNR family. 

\subsection{Visible Volume}
\label{sec:vvol_ul}

In general, the visible volume~\cite{Collaboration:2011fk} is a function of the component masses ($m_1$, $m_2$) of the binary system. It can be calculated as 
\begin{equation}
\label{eqn:intvv}
\VV(m_1,m_2,\eta) = 4\pi\int_0^\infty \epsilon(r, m_1, m_2, \eta) r^2 dr\,.
\end{equation}
where $\epsilon$ is the detection efficiency of the search, which is also a function of the distance to the source $r$. The visible volume is calculated for a given threshold on the coherent network amplitude $\eta$. To display the dependence on component masses, the visible volume is binned (25\msun$\times$25\msun bins) in the component mass plane. Here, we also assume that the detection efficiency is averaged over the sky position, binary inclination, and polarization angles.

Instead of a direct calculation of Eq.~\ref{eqn:intvv}, the integral can be estimated as a sum over the inverse density of the detected injections. Namely, each injection is assigned a density number $\rho_i$ and the analysis of each injected event is a statistical trial of whether or not that density is detected for a given threshold on $\eta$. The integral~\ref{eqn:intvv} becomes then a sum over detected injections 
\begin{equation}
\VV(m_1,m_2,\eta) = \sum_i \frac{1}{\rho_i} = 
\sum_i {4\pi r_i^2}\left(\frac{dN_{\textrm{inj}}}{dr}(r_i)\right)^{-1}\,,
\label{eqn:vvol_dens}
\end{equation}
where $r_i$ is the distance to the $i^{th}$ injection, and $dN_{\textrm{inj}}/dr$ is the radial density of simulated events. They are injected into a spherical volume with a fiducial radius of 2~Gpc with the density distribution linearly increasing in distance and optimized to reduce the statistical errors. To express the search sensitivity, below we also use the effective range $R_{\textrm{eff}}$, which is calculated as the radius of the visible volume. 

\subsection{False Alarm Rate Density and Event Significance}
\label{sec:fad}

The methods employed in previous searches for calculation of event significance compare foreground events to the expected background. Given a foreground event with the coherent amplitude $\eta$, its significance is determined by the false alarm rate 
\begin{equation}
\textrm{FAR}(\eta)=\frac{N(\eta)}{T_{\textrm{bkg}}},
\label{eqn:far}
\end{equation}
where $T_{\textrm{bkg}}$ is the accumulated live time for the corresponding background sample and $N(\eta)$ is the number of background events with strength greater than $\eta$. However, the IMBH search combines searches from two different detector networks. Therefore, the coherent network amplitudes calculated for different networks are not directly comparable and the networks may have significantly different sensitivities and background rates. To combine the results of multiple searches into a single measurement, the IMBH search employs a statistical procedure based on the false alarm density (FAD) rate~\cite{Pankow:2011aa} defined as
\begin{equation}
\textrm{FAD}(\eta) = \frac{1}{T_{\textrm{bkg}}}\left(\sum_{\eta_i>\eta} \frac{1}{\VVavg(\eta_i)}\right)\,.
\label{eqn:fad}
\end{equation}
Given an event with the coherent amplitude $\eta$, its FAD rate is calculated from the mass averaged visible volume $\VVavg$ as a function of the coherent network amplitude. The sum is performed over the background events from the corresponding network with $\eta_i>\eta$. The FAD estimates the number of background events expected in a given network's visible volume. Whereas the FAR statistic takes into account only the background rates, the FAD statistic also includes the sensitivity of a network to a population of expected GW sources~\cite{Klimenko:2011ab,Klimenko:2012aa}. It weights search networks by their overall sensitivity to the source population and their background rates, and therefore allows a direct comparison between disparate networks. The FAD statistic is then used to rank candidates in the combined search. More significant candidates have smaller FAD rates. To determine the event significance, its FAD rate is compared to the  time-volume product of the combined search (the overall search productivity $\nu$):
\begin{equation}
\nu(\textrm{FAD}) = \sum_k T_{\textrm{obs}}[k]\VV[k](\textrm{FAD})\,,
\label{eqn:productivity}
\end{equation}
where the sum is over the networks and $T_{\textrm{obs}}$ is the observation time of each network (listed in Table~\ref{tbl:livetime}). The product $\mu=\textrm{FAD}\cdot\nu(\textrm{FAD})$ is the mean number of events expected from the background Poisson process. The false alarm probability (FAP) is calculated as
\begin{equation}
\textrm{FAP}(N) = 1-\sum_{n=0}^{N-1}\frac{\mu^n}{n!}\exp(-\mu)\,.
\label{eqn:expevn}
\end{equation}
where N is the number of foreground events below a given FAD value. The FAP value indicates the probability that the candidate events are originating from a non-GW process.

\subsection{Statistical and Systematic Errors}
\label{sec:unc}

There are several uncertainties associated with the estimation of the visible volume such as statistical errors due to a limited number of simulated events, calibration errors of the detector data streams, and systematic errors due to uncertainties of the simulated IMBH waveforms arising from differences between the waveforms and nature. 

By using binomial statistics and Eq.~(\ref{eqn:vvol_dens}) the statistical uncertainty on the visible volume can be estimated as
\begin{equation}
\label{eqn:vvol_unc}
\delta\VV = \sqrt{\sum_i\frac{1}{\rho_i^2}} \,,
\end{equation}
where the sum is taken over the detected injections. The approximations used in the calculation could only increase the uncertainty, and therefore the estimate~(\ref{eqn:vvol_unc}) is conservative. The statistical error in any given component mass bin is usually less than 5\%.

The calibration procedure  of the GW strain data and the associated uncertainties for the S5/VSR1 run are described elsewhere~\cite{Collaboration:2010fl,collaboration:2010up}. The amplitude calibration error~\cite{LIGO:2009pz,Abadie:2010xk} directly translates into the error on the effective range of the search ($<$11\%). Respectively, the error on the visible volume is approximately 33\%.

Further checks were performed using an updated EOBNR (EOBNRv2)~\cite{Pan:2011fk} family which includes more PN corrections. Comparisons between the EOBNRv2 family, the IMRPhenom family, and waveforms drawn directly from numerical relativity simulations agree to within 15\% in the SNR induced in the detectors. Propagating this to the volume by noting that SNR is proportional to distance, we estimate a conservative systematic uncertainty on the search visible volume of 45\% due to imperfect knowledge of the IMBH waveforms used in the simulations.

The EOBNR waveform family (EOBNRv1)~\cite{Buonanno:2007sf} used to calculate the visible volume in this search predicts more GW power radiated during the inspiral and merger phase than seen in numerical relativity simulations. As a result, \emph{a priori} this model allows sources that are farther away to be seen. To account for the overestimation of the visible volume, the search for IMR signals in the S5 data between 25--100 \msun~\cite{Collaboration:2011fk} applied a distance correction. We follow the same procedure in the IMBH search. The newer EOBNRv2 family also includes higher order PN corrections to the dominant (2,2) mode and an improved calibration of the frequency evolution which manifest as a systematic shift in its noise weighted power as a function of frequency and a slight time dependence to the polarization. It is observed as an additional 10\% bias in terms of $\eta$ and a corresponding loss of detection efficiency due to the constant polarization constraint. For the case near the most sensitive region in the component mass space (centered around 88+88 \msun), the correction reduces the effective range of the search by about 50\%. The effect on larger total mass systems is even more pronounced.

It should be noted the EOBNR model used in the simulations neglects contributions to the waveform from other multipolar harmonics: $l,m$ modes different from the dominant 2,2 harmonic were not included in the initial EOBNR model. These modes could provide additionally detectable power. However, they interfere both constructively and destructively with the dominant harmonic and this may break the exact elliptical dependence (i.e. 90${}^{\circ}$ phase shift) between the two polarization states. Overall, the effect on the detected power by these modes is expected be small in regards to other sources of uncertainty~\cite{2010PhRvD..82j4006O}.

Propagating all the remaining uncertainties considered here into the volume gives an overall uncertainty of 60\% in volume. The rate density estimates are then readjusted upward accordingly by the same amount.

%% file: results.tex
\section{Results and Discussion}
\label{sec:results}

No event candidates were found to be significant to claim a detection. Therefore, we place upper limits on the rate density of IMBH coalescences as a function of the component masses.

\subsection{Event Candidates}

\begin{figure}[htbp]
\begin{center}
\includegraphics[scale=0.45,trim=0.0in 0.0in 0.0in 0.0in,clip]{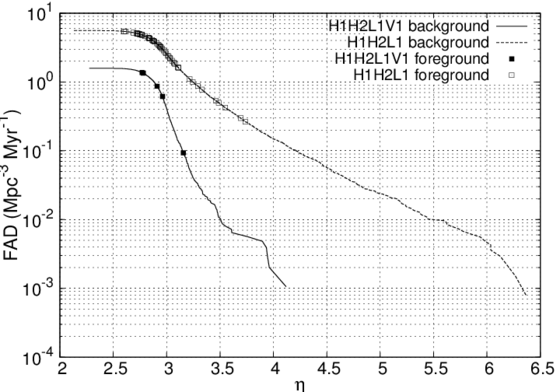}
\caption{False alarm density rate vs $\eta$ for the background events (H1H2L1V1 - solid line, H1H2L1 - dashed line) and the foreground events (H1H2L1V1 - black squares,  H1H2L1 - open squares).}
\label{fig:fad}
\end{center}
\end{figure}

The FAD rate distributions for the background and foreground events are shown in figure~\ref{fig:fad} as a function of the coherent network amplitude. All events are ranked by their FAD rate, with the most significant events represented by the low FAD values. Several foreground events with the lowest FAD rates are shown in Table~\ref{tbl:events}. The first (rank 1) event with the lowest FAD rate is produced by the four detector network with $\eta=3.16$. This event has an associated FAP of 45\%, which is not considered to be significant. No other event candidates have a low FAP sufficient for detection.

\begin{table*}[htbp]
\begin{center}
\caption{Highest ranked events by FAD. The first ranked event, produced by the four detector network, has a relatively small $\eta$ compared to the other three events. However, the four detector network is much less noisy resulting in a low FAD value.}
\begin{tabular}{ccccccccc}
\hline 
rank & GPS time & network & $\eta$ & cc & $\lambda_{\mathrm{net}}$ & FAR (yr${}^{-1}$) & FAD (Mpc${}^{-3}$Myr${}^{-1}$) & FAP \\
\hline \hline 
1 & 871474393 & H1H2L1V1 & 3.16 & 0.90 & 0.17 & 0.76 & 0.09 & 45\% \\
2 & 857692870 & H1H2L1   & 3.74 & 0.74 & 0.13 & 1.61 & 0.26 & 63\%\\
3 & 846735754 & H1H2L1   & 3.69 & 0.76 & 0.13 & 1.91 & 0.30 & 45\% \\
4 & 820091022 & H1H2L1   & 3.55 & 0.83 & 0.05 & 2.90 & 0.42 & 51\% \\
\hline
\end{tabular}
\label{tbl:events}
\end{center}
\end{table*}

\subsection{Visible Volume and Rate Limits}

\begin{figure*}[htbp]
\begin{center}
\includegraphics[scale=0.45]{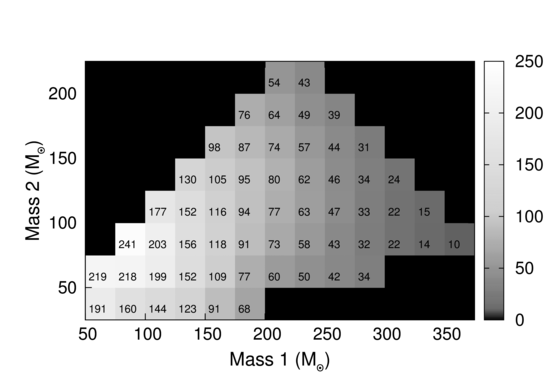} 
\includegraphics[scale=0.45]{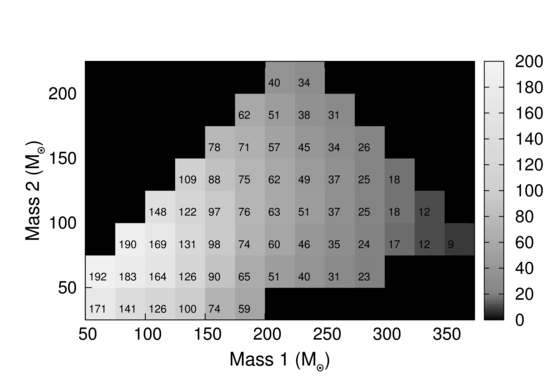} 
\caption{The effective range $R_{\textrm{eff}}$ in Mpc (values and white-gray scale) per component mass bin: H1H2L1V1 (left plot), H1H2L1 (right plot). The overall uncertainty on the ranges quoted here is 20\%. IMBH systems with mass ratios of greater than 4:1 are excluded since their waveforms are not verified with the NR calculations.}
\label{fig:visvol}
\end{center}
\end{figure*}

The visible volume is calculated from equation~\ref{eqn:vvol_dens} for the events binned in the component mass plane. The thresholds on the coherent amplitude of each network are determined by the FAD rate of the loudest event (see Figure~\ref{fig:fad}), denoted below in the text as $\textrm{FAD}^{\star}$. Figure~\ref{fig:visvol} shows the effective range as a function of the component masses for the networks analyzed in this search. The mass bins are limited to mass ratios less than 4:1, since no numerical relativity data is readily available for validation of the waveforms with larger mass ratios. For the more sensitive H1H2L1V1 network the best effective range is achieved in the 88+88~\msun bin at 241 Mpc. For the H1H2L1 network the corresponding range is 190 Mpc. The ranges in Figure~\ref{fig:visvol} take into account the SNR bias correction for the EOBNR waveform family as described in section \ref{sec:unc}. Combining errors from the statistical procedure, calibration and the waveform systematic errors, the total uncertainty on the effective ranges is estimated to be 20\%. 

\begin{figure}[htbp]
\begin{center}
\includegraphics[scale=0.45]{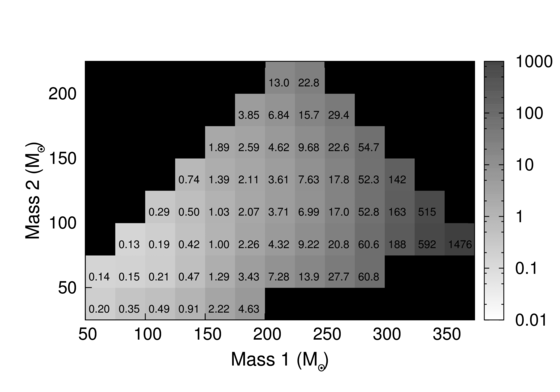} 
\caption{The rate density upper limits in Mpc${}^{-3}$Myr${}^{-1}$ (values and white-gray scale) calculated over the component mass plane. These rate limits include all sources of error.} 
\label{fig:rate_upper_limit}
\end{center}
\end{figure}

In the absence of detection, we set upper limits on the rate of IMBH mergers at the 90\% confidence level by using the loudest event statistic~\cite{Biswas:2007vn}:
\begin{equation}
\label{eqn:rate}
R_{90\%} = \frac{2.3}{\nu(\textrm{FAD}^{\star})}\,.
\end{equation}
The $\nu(\textrm{FAD}^{\star})$ is the time-volume productivity of the search calculated at the FAD rate of the first ranked event (the top event in Table \ref{tbl:events}). The rate density upper limits calculated in a binning of the component masses are presented in Figure \ref{fig:rate_upper_limit}. The upper limit for the combined search, averaged over all masses, is estimated to be 0.9~\MpcMyr. In the most sensitive bin, the rate limit is nearly an order of magnitude greater than for the overall search at 0.13~\MpcMyr. 

Since globular clusters are the most likely hosts of IMBHs, we convert our overall search upper limit into an astrophysical density of 3\e{3} \Gcyr. This rate is still few orders of magnitude above the predictions for IMBH-IMBH rates in~\cite{ratesdoc}. It should be noted, however, that the predicted astrophysical rates are very uncertain due to the lack of knowledge of the distribution and formation of IMBH sources.

%% file: summary.tex
\section{Summary}
\label{sec:sum}

In this paper we have presented a search for gravitational waves from IMBH coalescences in the mass range of 100--450 \msun and mass ratios up to 4:1. The search was performed in the S5/VSR1 data collected with two different network configurations: H1H2L1 and H1H2L1V1. For identification of potential GW candidates we used the Coherent WaveBurst algorithm with a polarization constraint for the first time. To establish the significance of candidates from either search network, we combined their detection statistics into a single measurement by using the false alarm rate density statistic. No plausible GW candidates were identified. From this search, we place upper limits on the rate density of the IMBH binaries as a function of the component masses. In the most sensitive mass bin (centered at 88+88~\msun) the rate limit is 0.13 \MpcMyr. When averaged over the mass plane the rate limit is 0.9 \MpcMyr at the 90\% confidence level. 

The sensitivity of the search was estimated by Monte Carlo simulations of detection efficiency using waveforms from the EOBNR family with component masses uniformly distributed on the mass plane. The most dominant source of error in this analysis is the systematic uncertainty (45\%) due to accuracy of simulated IMBH waveforms used for the estimation of the search visible volume. There are a few features of black hole coalescence which were neglected in the simulation studies, for instance, the effect of spinning component black holes. However, un-modeled searches like CWB are sensitive to the energy emitted in gravitational waves regardless of details of the waveform evolution.  While the effects of spinning component masses in the binary have not been examined in this study in detail, it is expected that their inclusion could only increase the effective range  of the search~\cite{Reisswig:2009fk}. This is because the increase in the GW energy output in favorable (co-aligned) spin configurations is greater than its decrease from less favorable (anti-aligned) configurations, giving an overall increase in the emitted  energy. Moreover, the additional energy of aligned spin configurations could extend the mass range for which we can search beyond 450 \msun.

This search has been limited to a relatively small area in the component mass plane of potential IMBH sources. However, future experiments with advanced detectors will have a significant increase in sensitivity, and more importantly, advanced detectors will also widen the sensitive frequency band. At low frequencies, LIGO design sensitivity at 10 Hz calls for an increase of a few orders of magnitude; hence a greater fraction of IMBH binary signals should become observable. These improvements should allow for better chances of detection of IMBH sources.

%% file: acknowledgements.tex
\begin{acknowledgments}

The authors gratefully acknowledge the support of the United States National 
Science Foundation for the construction and operation of the LIGO Laboratory, 
the Science and Technology Facilities Council of the United Kingdom, the 
Max-Planck-Society and the State of Niedersachsen/Germany for support of the 
construction and operation of the GEO\,600 detector, 
and the Italian Istituto Nazionale di Fisica Nucleare and the French Centre 
National de la Recherche Scientifique for the construction and operation of 
the Virgo detector. The authors also gratefully acknowledge the support of 
the research by these agencies and by the Australian Research Council, the 
Council of Scientific and Industrial Research of India, the Istituto 
Nazionale di Fisica Nucleare of Italy, the Spanish Ministerio de Educaci\'on 
y Ciencia, the Conselleria d'Economia Hisenda i Innovaci\'o of the Govern de 
les Illes Balears, the Foundation for Fundamental Research on Matter supported 
by the Netherlands Organisation for Scientific Research,
the Polish Ministry of Science and Higher Education, the FOCUS Programme
of Foundation for Polish Science, the Royal Society, 
the Scottish Funding Council, the Scottish Universities Physics Alliance, 
the National Aeronautics and Space Administration, the Carnegie Trust, 
the Leverhulme Trust, the David and Lucile Packard Foundation, the Research 
Corporation, and the Alfred P. Sloan Foundation. 
This document has been assigned LIGO Laboratory document number P1100068.

\end{acknowledgments}

%% file: appendix_ner.tex
\section{Likelihood Analysis of Elliptically Polarized Waves}
\label{elliptical}

The CWB algorithm performs the likelihood analysis~\cite{Klimenko:2005wa} of the detector data streams, which are transformed into the time-frequency domain with the Meyer wavelet~\cite{Vidakovic:1999fk}. The sampled network data is ${\bf{x}}[i]=[x_1[i],...,x_K[i]]$, where $k$ is an index iterating over the $K$ detectors in the network. The vector ${\bf{x}}[i]$ is a function of the time and frequency indicated with a single time-frequency index $i$, which is often omitted later in the text. The likelihood ratio is defined as
\begin{equation}
\label{eq:LIKE}
\Lambda({{\bf{x}},\Omega}) = \frac{p({\bf{x}}|{\bf{h}}(\Omega))}{p({\bf{x}}|0)}\;,
\end{equation}
where $\Omega$ is a parameter set describing two GW polarizations ${\bf{h}}=(h_+,h_\times)$, $p({\rm x}|0)$ is the joint probability that the data is only instrumental noise, and $p({\rm x}|{\bf{h}})$ is the joint probability that a GW signal ${\bf{h}}$ is present in the data. The explicit form of the likelihood ratio is determined by the noise model $p({\bf{x}}|0)$ and by the signal model ${\bf{h}}(\Omega)$. In the analysis we assume that the noise of detectors is Gaussian with the standard deviations $\sigma_k[i]$. To account for the time and frequency variability of the noise, the $\sigma_k[i]$ are estimated for every time-frequency sample. For elliptically polarized waves originating at a sky location ($\theta$,$\phi$), instead of reconstructing two unknown signal polarizations $h_+[i]$ and $h_\times[i]$ (un-modeled case), only one waveform $h[i]$ and two other signal parameters need to be reconstructed: the ellipticity parameter $\alpha$ (related to the inclination angle of the binary axis) and the polarization angle $\Psi$. Therefore the signal model is introduced into the analysis by the following parameterization of the detector response
\begin{align}
\label{hdet0}
\bfxi_{\mathrm{h}}[i] &= {\bf{f}}_{+}(\Omega,\Psi) h[i] + \alpha {\bf{f}}_{\times}(\Omega,\Psi) \tilde{h}[i] \;, \\
\label{hdet90}
\tilde\bfxi_{\mathrm{h}}[i] &= {\bf{f}}_{+}(\Omega,\Psi) \tilde{h}[i] - \alpha {\bf{f}}_{\times}(\Omega,\Psi) h[i] \;, 
\end{align}
where $\tilde{h}$ and $\tilde\bfxi_{\mathrm{h}}$ are the $90^o$ phase shifted counterparts of ${h}$ and $\bfxi_{\mathrm{h}}$. The components of the response vector $\bfxi_{\mathrm{h}}[i]$, the noise scaled antenna pattern vectors ${\bf{f}}_{+(\times)}[i]$ and the noise scaled network data vector ${\bf{w}}[i]$ represent the individual detectors:
\begin{align}
\label{hfV}
{\bfxi_{\mathrm{h}}}[i]&=\left(\xi_{\mathrm{h}1}[i],...,\xi_{\mathrm{h}K}[i]\right), \\
{\bf{f_{+(\times)}}} &= \left(\frac{f_{1+(\times)}}{\sigma_1[i]},..,\frac{f_{K+(\times)}}{\sigma_K[i]} \right), \\
{\bf{w}}[i]&=\left(\frac{x_1[i]}{\sigma_1[i]},...,\frac{x_K[i]}{\sigma_K[i]}\right) \;.
\end{align}
The solutions for $h$, $\alpha$, and $\Psi$ are obtained by variation of the
combined likelihood functional 
${\cal{L}}({\bf{w}}|{\bfxi_{\mathrm{h}}})$+${{\cal{L}}(\tilde{\bf{w}}|\tilde{\bfxi}_{\mathrm{h}}})$, 
where $\tilde{\bf{w}}$ is the quadrature of ${\bf{w}}$. The likelihood functional ${\cal{L}}({\bf{w}}|\bfxi_{\mathrm{h}})$ is defined as twice the logarithm of $\Lambda$
\begin{equation}
\label{eq:like}
{\cal{L}}({\bf{w}}|{\bfxi_{\mathrm{h}}}) = 2({\bf{w}} \vert \bfxi_{\text{h}})- (\bfxi_{\text{h}} \vert \bfxi_{\text{h}}) \, ,
\end{equation}
where the inner products are given by a sum over a time-frequency area $I$ containing the signal
\begin{equation}
\label{eq:inner}
({\bf{a}}|{\bf{b}}) = \sum_{i\in{I}}{({\bf{a}}[i] \cdot {\bf{b}}[i])} \;. 
\end{equation}
The reconstructed network responses $\xi[i]$ and $\tilde{\xi}[i]$  are obtained by substituting the solutions for $h$, $\alpha$ and $\Psi$ into the equations~\ref{hdet0} and~\ref{hdet90}. Respectively the reconstructed null stream is ${\bf{n}}[i]={\bf{w}}[i]-\bfxi[i]$. 

The three data streams ${\bf{w}}$, $\bfxi$ and ${\bf{n}}$ are used to calculate the normalized (by the noise variance) energies of the network data stream $E_{\textrm{tot}}$, the signal energy  $E_{\textrm{GW}}$ and the noise energy $E_n$ respectively. The $E_{\textrm{GW}}=({\bfxi}|{\bfxi})=\sum_{i,j}{L_{ij}}$, where the components of the likelihood matrix $L_{ij}$ are calculated from the output of the corresponding detector pair ($i,j$). The sum of the off-diagonal terms ($i \ne j$) of the matrix $L_{ij}$ define the network coherent energy $E_c=\sum_{i \ne j}{L_{ij}}$. The phase shifted coherent energy $\tilde{E}_c$ is similarly defined for the phase shifted likelihood. There are two coherent statistics, that are used for CWB selection cuts: the network correlation coefficient 
\begin{equation}
\textrm{cc} = \frac{E_c+\tilde{E}_c}{|E_c+\tilde{E}_c|+E_n+\tilde{E}_n},
\end{equation}
and the network energy disbalance  $\lambda_{\mathrm{net}} = \max(\lambda, \tilde{\lambda})$, where 
\begin{equation}
\lambda = \frac{1}{E_c}\sum_{k=1}^K{
	\left|\sum_{i\in{I}} {\xi_k[i] n_k[i] } \right|
}
\end{equation}
and $\tilde{E}_c$, $\tilde{E}_n$, and $\tilde{\lambda}$ are calculated for the $90^o$ phase shifted data streams. The CWB algorithm also defines the reduced coherent energy
\begin{equation}
e_c = \sum_{i \ne j}{\frac{L^2_{ij}}{\sqrt{L_{ii} L_{jj}}}},
\end{equation}
which is used for calculation of the main CWB detection statistic --- the coherent network amplitude
\begin{equation}
\eta = \sqrt{\frac{e_c\,\textrm{cc}}{K}}.
\end{equation}
It is used to rank CWB events and establish their significance against a sample of background events.

%% file: bbhpaper.bbl
\begin{thebibliography}{63}
\expandafter\ifx\csname natexlab\endcsname\relax\def\natexlab#1{#1}\fi
\expandafter\ifx\csname bibnamefont\endcsname\relax
  \def\bibnamefont#1{#1}\fi
\expandafter\ifx\csname bibfnamefont\endcsname\relax
  \def\bibfnamefont#1{#1}\fi
\expandafter\ifx\csname citenamefont\endcsname\relax
  \def\citenamefont#1{#1}\fi
\expandafter\ifx\csname url\endcsname\relax
  \def\url#1{\texttt{#1}}\fi
\expandafter\ifx\csname urlprefix\endcsname\relax\def\urlprefix{URL }\fi
\providecommand{\bibinfo}[2]{#2}
\providecommand{\eprint}[2][]{\url{#2}}

\bibitem[{\citenamefont{Abbott
  et~al.}(2009{\natexlab{a}})}]{0034-4885-72-7-076901}
\bibinfo{author}{\bibfnamefont{B.~P.} \bibnamefont{Abbott}}
  \bibnamefont{et~al.}, \bibinfo{journal}{Rep. Prog. Phys.}
  \textbf{\bibinfo{volume}{72}}, \bibinfo{pages}{076901}
  (\bibinfo{year}{2009}{\natexlab{a}}).

\bibitem[{\citenamefont{Acernese et~al.}(2006)}]{0264-9381-23-19-S01}
\bibinfo{author}{\bibfnamefont{F.}~\bibnamefont{Acernese}}
  \bibnamefont{et~al.}, \bibinfo{journal}{Class. Quant. Grav.}
  \textbf{\bibinfo{volume}{23}}, \bibinfo{pages}{S635} (\bibinfo{year}{2006}).

\bibitem[{\citenamefont{{Grote} and {the LIGO Scientific
  Collaboration}}(2008)}]{Grote:2008}
\bibinfo{author}{\bibfnamefont{H.}~\bibnamefont{{Grote}}} \bibnamefont{and}
  \bibinfo{author}{\bibnamefont{{the LIGO Scientific Collaboration}}},
  \bibinfo{journal}{Class. Quant. Grav.} \textbf{\bibinfo{volume}{25}},
  \bibinfo{pages}{114043} (\bibinfo{year}{2008}).

\bibitem[{\citenamefont{Acernese et~al.}(2008)}]{Acernese:2008kx}
\bibinfo{author}{\bibfnamefont{F.}~\bibnamefont{Acernese}}
  \bibnamefont{et~al.}, \bibinfo{journal}{J. Opt. A: Pure Appl. Opt.}
  \textbf{\bibinfo{volume}{10}}, \bibinfo{pages}{064009}
  (\bibinfo{year}{2008}).

\bibitem[{\citenamefont{Blanchet}(2001)}]{Blanchet:2001og}
\bibinfo{author}{\bibfnamefont{L.}~\bibnamefont{Blanchet}}, in
  \emph{\bibinfo{booktitle}{Gravitatonal Waves, Proceedings of the Como School
  on Gravitational Waves in Astrophysics}}, edited by
  \bibinfo{editor}{\bibfnamefont{I.}~\bibnamefont{Ciufolini}},
  \bibinfo{editor}{\bibfnamefont{V.}~\bibnamefont{Gorini}},
  \bibinfo{editor}{\bibfnamefont{U.}~\bibnamefont{Moschella}},
  \bibnamefont{and} \bibinfo{editor}{\bibfnamefont{P.}~\bibnamefont{Fre}}
  (\bibinfo{publisher}{Institute of Physics Publishing}, \bibinfo{year}{2001}).

\bibitem[{\citenamefont{Blanchet}(2002)}]{Blanchet:2002av}
\bibinfo{author}{\bibfnamefont{L.}~\bibnamefont{Blanchet}},
  \bibinfo{journal}{Living Rev. Rel.} \textbf{\bibinfo{volume}{5}},
  \bibinfo{pages}{3} (\bibinfo{year}{2002}).

\bibitem[{\citenamefont{Boyle et~al.}(2007)}]{Boyle:2007ft}
\bibinfo{author}{\bibfnamefont{M.}~\bibnamefont{Boyle}} \bibnamefont{et~al.},
  \bibinfo{journal}{Phys.~Rev.~D} \textbf{\bibinfo{volume}{76}},
  \bibinfo{pages}{124038} (\bibinfo{year}{2007}).

\bibitem[{\citenamefont{Blanchet et~al.}(1995)\citenamefont{Blanchet, Damour,
  and Iyer}}]{PhysRevD.51.5360}
\bibinfo{author}{\bibfnamefont{L.}~\bibnamefont{Blanchet}},
  \bibinfo{author}{\bibfnamefont{T.}~\bibnamefont{Damour}}, \bibnamefont{and}
  \bibinfo{author}{\bibfnamefont{B.~R.} \bibnamefont{Iyer}},
  \bibinfo{journal}{Phys. Rev. D} \textbf{\bibinfo{volume}{51}},
  \bibinfo{pages}{5360} (\bibinfo{year}{1995}).

\bibitem[{\citenamefont{Buonanno and Damour}(1999)}]{Buonanno:1999sp}
\bibinfo{author}{\bibfnamefont{A.}~\bibnamefont{Buonanno}} \bibnamefont{and}
  \bibinfo{author}{\bibfnamefont{T.}~\bibnamefont{Damour}},
  \bibinfo{journal}{Phys. Rev. D} \textbf{\bibinfo{volume}{59}},
  \bibinfo{pages}{084006} (\bibinfo{year}{1999}).

\bibitem[{\citenamefont{Damour et~al.}(2008)\citenamefont{Damour, Nagar,
  Dorband, Pollney, and Rezzolla}}]{Damour:2008lu}
\bibinfo{author}{\bibfnamefont{T.}~\bibnamefont{Damour}},
  \bibinfo{author}{\bibfnamefont{A.}~\bibnamefont{Nagar}},
  \bibinfo{author}{\bibfnamefont{E.~N.} \bibnamefont{Dorband}},
  \bibinfo{author}{\bibfnamefont{D.}~\bibnamefont{Pollney}}, \bibnamefont{and}
  \bibinfo{author}{\bibfnamefont{L.}~\bibnamefont{Rezzolla}},
  \bibinfo{journal}{Phys. Rev. D} \textbf{\bibinfo{volume}{77}},
  \bibinfo{pages}{084017} (\bibinfo{year}{2008}).

\bibitem[{\citenamefont{Ajith et~al.}(2008)}]{Ajith:2007fj}
\bibinfo{author}{\bibfnamefont{P.}~\bibnamefont{Ajith}} \bibnamefont{et~al.},
  \bibinfo{journal}{Phys. Rev. D} \textbf{\bibinfo{volume}{77}},
  \bibinfo{pages}{104017} (\bibinfo{year}{2008}).

\bibitem[{\citenamefont{Buonanno
  et~al.}(2007{\natexlab{a}})\citenamefont{Buonanno, Pan, Baker, Centrella,
  Kelly, McWilliams, and van Meter}}]{Buonanno:2007sf}
\bibinfo{author}{\bibfnamefont{A.}~\bibnamefont{Buonanno}},
  \bibinfo{author}{\bibfnamefont{Y.}~\bibnamefont{Pan}},
  \bibinfo{author}{\bibfnamefont{J.~G.} \bibnamefont{Baker}},
  \bibinfo{author}{\bibfnamefont{J.}~\bibnamefont{Centrella}},
  \bibinfo{author}{\bibfnamefont{B.~J.} \bibnamefont{Kelly}},
  \bibinfo{author}{\bibfnamefont{S.~T.} \bibnamefont{McWilliams}},
  \bibnamefont{and} \bibinfo{author}{\bibfnamefont{J.~R.} \bibnamefont{van
  Meter}}, \bibinfo{journal}{Phys. Rev. D} \textbf{\bibinfo{volume}{76}},
  \bibinfo{pages}{104049} (\bibinfo{year}{2007}{\natexlab{a}}).

\bibitem[{\citenamefont{Buonanno
  et~al.}(2009{\natexlab{a}})\citenamefont{Buonanno, Iyer, Ochsner, Pan, and
  Sathyaprakash}}]{Buonanno:2009sf}
\bibinfo{author}{\bibfnamefont{A.}~\bibnamefont{Buonanno}},
  \bibinfo{author}{\bibfnamefont{B.~R.} \bibnamefont{Iyer}},
  \bibinfo{author}{\bibfnamefont{E.}~\bibnamefont{Ochsner}},
  \bibinfo{author}{\bibfnamefont{Y.}~\bibnamefont{Pan}}, \bibnamefont{and}
  \bibinfo{author}{\bibfnamefont{B.~S.} \bibnamefont{Sathyaprakash}},
  \bibinfo{journal}{Phys. Rev. D} \textbf{\bibinfo{volume}{80}},
  \bibinfo{pages}{084043} (\bibinfo{year}{2009}{\natexlab{a}}).

\bibitem[{\citenamefont{Buonanno
  et~al.}(2007{\natexlab{b}})\citenamefont{Buonanno, Cook, and
  Pretorius}}]{Buonanno:2007qm}
\bibinfo{author}{\bibfnamefont{A.}~\bibnamefont{Buonanno}},
  \bibinfo{author}{\bibfnamefont{G.~B.} \bibnamefont{Cook}}, \bibnamefont{and}
  \bibinfo{author}{\bibfnamefont{F.}~\bibnamefont{Pretorius}},
  \bibinfo{journal}{Phys. Rev. D} \textbf{\bibinfo{volume}{75}},
  \bibinfo{pages}{124018} (\bibinfo{year}{2007}{\natexlab{b}}).

\bibitem[{\citenamefont{Abbott et~al.}(2008{\natexlab{a}})}]{Abbot:2007uq}
\bibinfo{author}{\bibfnamefont{B.}~\bibnamefont{Abbott}} \bibnamefont{et~al.},
  \bibinfo{journal}{Phys. Rev. D} \textbf{\bibinfo{volume}{77}},
  \bibinfo{pages}{062002} (\bibinfo{year}{2008}{\natexlab{a}}).

\bibitem[{\citenamefont{Abbott et~al.}(2009{\natexlab{b}})}]{Abbott:2009bh}
\bibinfo{author}{\bibfnamefont{B.}~\bibnamefont{Abbott}} \bibnamefont{et~al.},
  \bibinfo{journal}{Phys. Rev. D} \textbf{\bibinfo{volume}{79}},
  \bibinfo{pages}{122001} (\bibinfo{year}{2009}{\natexlab{b}}).

\bibitem[{\citenamefont{Abbott
  et~al.}(2009{\natexlab{c}})}]{Collaboration:2009fy}
\bibinfo{author}{\bibfnamefont{B.}~\bibnamefont{Abbott}} \bibnamefont{et~al.},
  \bibinfo{journal}{Phys. Rev. D} \textbf{\bibinfo{volume}{80}},
  \bibinfo{pages}{047101} (\bibinfo{year}{2009}{\natexlab{c}}).

\bibitem[{\citenamefont{Abadie et~al.}(2011)}]{Collaboration:2011fk}
\bibinfo{author}{\bibfnamefont{J.}~\bibnamefont{Abadie}} \bibnamefont{et~al.},
  \bibinfo{journal}{Phys. Rev. D} \textbf{\bibinfo{volume}{83}},
  \bibinfo{pages}{122005} (\bibinfo{year}{2011}).

\bibitem[{\citenamefont{Abbott et~al.}(2008{\natexlab{b}})}]{Abbott:2008eh}
\bibinfo{author}{\bibfnamefont{B.}~\bibnamefont{Abbott}} \bibnamefont{et~al.},
  \bibinfo{journal}{Class. Quant. Grav.} \textbf{\bibinfo{volume}{25}},
  \bibinfo{pages}{245008} (\bibinfo{year}{2008}{\natexlab{b}}).

\bibitem[{\citenamefont{Abbott et~al.}(2009{\natexlab{d}})}]{LIGO:2009pz}
\bibinfo{author}{\bibfnamefont{B.}~\bibnamefont{Abbott}} \bibnamefont{et~al.},
  \bibinfo{journal}{Phys. Rev. D} \textbf{\bibinfo{volume}{80}}
  (\bibinfo{year}{2009}{\natexlab{d}}).

\bibitem[{\citenamefont{Abadie et~al.}(2010{\natexlab{a}})}]{Abadie:2010xk}
\bibinfo{author}{\bibfnamefont{J.}~\bibnamefont{Abadie}} \bibnamefont{et~al.},
  \bibinfo{journal}{Phys. Rev. D} \textbf{\bibinfo{volume}{81}},
  \bibinfo{pages}{102001} (\bibinfo{year}{2010}{\natexlab{a}}).

\bibitem[{\citenamefont{Pankow et~al.}(2009)\citenamefont{Pankow, Klimenko,
  Mitselmakher, Yakushin, Vedovato, Drago, Mercer, and Ajith}}]{Pankow:2009lv}
\bibinfo{author}{\bibfnamefont{C.}~\bibnamefont{Pankow}},
  \bibinfo{author}{\bibfnamefont{S.}~\bibnamefont{Klimenko}},
  \bibinfo{author}{\bibfnamefont{G.}~\bibnamefont{Mitselmakher}},
  \bibinfo{author}{\bibfnamefont{I.}~\bibnamefont{Yakushin}},
  \bibinfo{author}{\bibfnamefont{G.}~\bibnamefont{Vedovato}},
  \bibinfo{author}{\bibfnamefont{M.}~\bibnamefont{Drago}},
  \bibinfo{author}{\bibfnamefont{R.~A.} \bibnamefont{Mercer}},
  \bibnamefont{and} \bibinfo{author}{\bibfnamefont{P.}~\bibnamefont{Ajith}},
  \bibinfo{journal}{Class. Quant. Grav.} \textbf{\bibinfo{volume}{26}},
  \bibinfo{pages}{204004} (\bibinfo{year}{2009}).

\bibitem[{\citenamefont{{S. F. Portegies Zwart} and
  McMillan}(2002)}]{0004-637X-576-2-899}
\bibinfo{author}{\bibnamefont{{S. F. Portegies Zwart}}} \bibnamefont{and}
  \bibinfo{author}{\bibfnamefont{S.~L.~W.} \bibnamefont{McMillan}},
  \bibinfo{journal}{Astrophys. J.} \textbf{\bibinfo{volume}{576}},
  \bibinfo{pages}{899} (\bibinfo{year}{2002}).

\bibitem[{\citenamefont{{Miller} and {Colbert}}(2004)}]{2004IJMPD..13....1M}
\bibinfo{author}{\bibfnamefont{M.~C.} \bibnamefont{{Miller}}} \bibnamefont{and}
  \bibinfo{author}{\bibfnamefont{E.~J.~M.} \bibnamefont{{Colbert}}},
  \bibinfo{journal}{Int. J. Mod. Phys. D} \textbf{\bibinfo{volume}{13}},
  \bibinfo{pages}{1} (\bibinfo{year}{2004}).

\bibitem[{\citenamefont{Ohkubo et~al.}(2009)\citenamefont{Ohkubo, Nomoto,
  Umeda, Yoshida, and Tsuruta}}]{0004-637X-706-2-1184}
\bibinfo{author}{\bibfnamefont{T.}~\bibnamefont{Ohkubo}},
  \bibinfo{author}{\bibfnamefont{K.}~\bibnamefont{Nomoto}},
  \bibinfo{author}{\bibfnamefont{H.}~\bibnamefont{Umeda}},
  \bibinfo{author}{\bibfnamefont{N.}~\bibnamefont{Yoshida}}, \bibnamefont{and}
  \bibinfo{author}{\bibfnamefont{S.}~\bibnamefont{Tsuruta}},
  \bibinfo{journal}{Astrophys. J.} \textbf{\bibinfo{volume}{706}},
  \bibinfo{pages}{1184} (\bibinfo{year}{2009}).

\bibitem[{\citenamefont{Miller and Hamilton}(2002)}]{Miller:2002uq}
\bibinfo{author}{\bibfnamefont{M.~C.} \bibnamefont{Miller}} \bibnamefont{and}
  \bibinfo{author}{\bibfnamefont{D.~P.} \bibnamefont{Hamilton}},
  \bibinfo{journal}{Mon. Not. Roy. Astron. Soc.} \textbf{\bibinfo{volume}{330}}
  (\bibinfo{year}{2002}).

\bibitem[{\citenamefont{Glebbeek et~al.}(2009)\citenamefont{Glebbeek, Gaburov,
  de~Mink, Pols, and {S. F. Portegies Zwart}}}]{refId}
\bibinfo{author}{\bibfnamefont{E.}~\bibnamefont{Glebbeek}},
  \bibinfo{author}{\bibfnamefont{E.}~\bibnamefont{Gaburov}},
  \bibinfo{author}{\bibfnamefont{S.~E.} \bibnamefont{de~Mink}},
  \bibinfo{author}{\bibfnamefont{O.~R.} \bibnamefont{Pols}}, \bibnamefont{and}
  \bibinfo{author}{\bibnamefont{{S. F. Portegies Zwart}}},
  \bibinfo{journal}{Astron. Astrophy.} \textbf{\bibinfo{volume}{497}},
  \bibinfo{pages}{255} (\bibinfo{year}{2009}).

\bibitem[{\citenamefont{Baumgardt et~al.}(2004)\citenamefont{Baumgardt, Makino,
  and Ebisuzaki}}]{0004-637X-613-2-1143}
\bibinfo{author}{\bibfnamefont{H.}~\bibnamefont{Baumgardt}},
  \bibinfo{author}{\bibfnamefont{J.}~\bibnamefont{Makino}}, \bibnamefont{and}
  \bibinfo{author}{\bibfnamefont{T.}~\bibnamefont{Ebisuzaki}},
  \bibinfo{journal}{Astrophys. J.} \textbf{\bibinfo{volume}{613}},
  \bibinfo{pages}{1143} (\bibinfo{year}{2004}).

\bibitem[{\citenamefont{O'Leary et~al.}(2006)\citenamefont{O'Leary, Rasio,
  Fregeau, Ivanova, and O'Shaughnessy}}]{0004-637X-637-2-937}
\bibinfo{author}{\bibfnamefont{R.~M.} \bibnamefont{O'Leary}},
  \bibinfo{author}{\bibfnamefont{F.~A.} \bibnamefont{Rasio}},
  \bibinfo{author}{\bibfnamefont{J.~M.} \bibnamefont{Fregeau}},
  \bibinfo{author}{\bibfnamefont{N.}~\bibnamefont{Ivanova}}, \bibnamefont{and}
  \bibinfo{author}{\bibfnamefont{R.}~\bibnamefont{O'Shaughnessy}},
  \bibinfo{journal}{Astrophys. J.} \textbf{\bibinfo{volume}{637}},
  \bibinfo{pages}{937} (\bibinfo{year}{2006}).

\bibitem[{\citenamefont{Strohmayer and Mushotzky}(2009)}]{Strohmayer:2009of}
\bibinfo{author}{\bibfnamefont{T.~E.} \bibnamefont{Strohmayer}}
  \bibnamefont{and} \bibinfo{author}{\bibfnamefont{R.~F.}
  \bibnamefont{Mushotzky}}, \bibinfo{journal}{Astrophys. J.}
  \textbf{\bibinfo{volume}{703}}, \bibinfo{pages}{1386} (\bibinfo{year}{2009}).

\bibitem[{\citenamefont{Colbert and Miller}(2004)}]{Colbert:2004sz}
\bibinfo{author}{\bibfnamefont{E.~J.~M.} \bibnamefont{Colbert}}
  \bibnamefont{and} \bibinfo{author}{\bibfnamefont{M.~C.}
  \bibnamefont{Miller}}, in \emph{\bibinfo{booktitle}{Tenth Marcel Grossman
  Meeting on General Relativity}}, edited by
  \bibinfo{editor}{\bibfnamefont{S.~P.-B.} \bibnamefont{M.~Novello}}
  \bibnamefont{and} \bibinfo{editor}{\bibfnamefont{R.}~\bibnamefont{Ruffini}}
  (\bibinfo{publisher}{World Scientific}, \bibinfo{year}{2004}).

\bibitem[{\citenamefont{Zampieri et~al.}(2010)\citenamefont{Zampieri, Colpi,
  Mapelli, Patruno, and Roberts}}]{Zampieri:2010uq}
\bibinfo{author}{\bibfnamefont{L.}~\bibnamefont{Zampieri}},
  \bibinfo{author}{\bibfnamefont{M.}~\bibnamefont{Colpi}},
  \bibinfo{author}{\bibfnamefont{M.}~\bibnamefont{Mapelli}},
  \bibinfo{author}{\bibfnamefont{A.}~\bibnamefont{Patruno}}, \bibnamefont{and}
  \bibinfo{author}{\bibfnamefont{T.~P.} \bibnamefont{Roberts}}, in
  \emph{\bibinfo{booktitle}{X-Ray Astronomy 2009: Present Status,
  Multiwavelength Approach and Future Perspectives}}, edited by
  \bibinfo{editor}{\bibfnamefont{L.~A.} \bibnamefont{A.~Comastri},
  \bibfnamefont{M.~Cappi}} (\bibinfo{publisher}{AIP}, \bibinfo{year}{2010}).

\bibitem[{\citenamefont{Okajima et~al.}(2006)\citenamefont{Okajima, Ebisawa,
  and Kawaguchi}}]{1538-4357-652-2-L105}
\bibinfo{author}{\bibfnamefont{T.}~\bibnamefont{Okajima}},
  \bibinfo{author}{\bibfnamefont{K.}~\bibnamefont{Ebisawa}}, \bibnamefont{and}
  \bibinfo{author}{\bibfnamefont{T.}~\bibnamefont{Kawaguchi}},
  \bibinfo{journal}{Astrophys. J. Lett.} \textbf{\bibinfo{volume}{652}},
  \bibinfo{pages}{L105} (\bibinfo{year}{2006}).

\bibitem[{\citenamefont{Casella et~al.}(2008)\citenamefont{Casella, Ponti,
  Patruno, Belloni, Miniutti, and Zampieri}}]{Casella:2008fk}
\bibinfo{author}{\bibfnamefont{P.}~\bibnamefont{Casella}},
  \bibinfo{author}{\bibfnamefont{G.}~\bibnamefont{Ponti}},
  \bibinfo{author}{\bibfnamefont{A.}~\bibnamefont{Patruno}},
  \bibinfo{author}{\bibfnamefont{T.}~\bibnamefont{Belloni}},
  \bibinfo{author}{\bibfnamefont{G.}~\bibnamefont{Miniutti}}, \bibnamefont{and}
  \bibinfo{author}{\bibfnamefont{L.}~\bibnamefont{Zampieri}},
  \bibinfo{journal}{Mon. Not. Roy. Astron. Soc.}
  \textbf{\bibinfo{volume}{287}}, \bibinfo{pages}{1707} (\bibinfo{year}{2008}).

\bibitem[{\citenamefont{Patruno and Zampieri}(2010)}]{Patruno:2010uq}
\bibinfo{author}{\bibfnamefont{A.}~\bibnamefont{Patruno}} \bibnamefont{and}
  \bibinfo{author}{\bibfnamefont{L.}~\bibnamefont{Zampieri}},
  \bibinfo{journal}{Mon. Not. Roy. Astron. Soc.}
  \textbf{\bibinfo{volume}{403}}, \bibinfo{pages}{L69} (\bibinfo{year}{2010}).

\bibitem[{\citenamefont{Baumgardt et~al.}(2005)\citenamefont{Baumgardt, Makino,
  and Hut}}]{0004-637X-620-1-238}
\bibinfo{author}{\bibfnamefont{H.}~\bibnamefont{Baumgardt}},
  \bibinfo{author}{\bibfnamefont{J.}~\bibnamefont{Makino}}, \bibnamefont{and}
  \bibinfo{author}{\bibfnamefont{P.}~\bibnamefont{Hut}},
  \bibinfo{journal}{Astrophys. J.} \textbf{\bibinfo{volume}{620}},
  \bibinfo{pages}{238} (\bibinfo{year}{2005}).

\bibitem[{\citenamefont{Safonova and Shastri}(2010)}]{Safonova:2009qp}
\bibinfo{author}{\bibfnamefont{M.}~\bibnamefont{Safonova}} \bibnamefont{and}
  \bibinfo{author}{\bibfnamefont{P.}~\bibnamefont{Shastri}},
  \bibinfo{journal}{Astrophysics and Space Science}
  \textbf{\bibinfo{volume}{325}}, \bibinfo{pages}{47} (\bibinfo{year}{2010}).

\bibitem[{\citenamefont{Vesperini et~al.}(2010)\citenamefont{Vesperini,
  McMillan, D'Ercole, and D'Antona}}]{Vesperini:2010kx}
\bibinfo{author}{\bibfnamefont{E.}~\bibnamefont{Vesperini}},
  \bibinfo{author}{\bibfnamefont{S.~L.} \bibnamefont{McMillan}},
  \bibinfo{author}{\bibfnamefont{A.}~\bibnamefont{D'Ercole}}, \bibnamefont{and}
  \bibinfo{author}{\bibfnamefont{F.}~\bibnamefont{D'Antona}},
  \bibinfo{journal}{Astrophys. J. Lett.} \textbf{\bibinfo{volume}{713}},
  \bibinfo{pages}{L41} (\bibinfo{year}{2010}).

\bibitem[{\citenamefont{G{\"u}ltekin et~al.}(2004)\citenamefont{G{\"u}ltekin,
  Miller, and Hamilton}}]{0004-637X-616-1-221}
\bibinfo{author}{\bibfnamefont{K.}~\bibnamefont{G{\"u}ltekin}},
  \bibinfo{author}{\bibfnamefont{M.~C.} \bibnamefont{Miller}},
  \bibnamefont{and} \bibinfo{author}{\bibfnamefont{D.~P.}
  \bibnamefont{Hamilton}}, \bibinfo{journal}{Astrophys. J.}
  \textbf{\bibinfo{volume}{616}}, \bibinfo{pages}{221} (\bibinfo{year}{2004}).

\bibitem[{\citenamefont{Umbreit et~al.}(2010)\citenamefont{Umbreit, Fregeau,
  and Rasio}}]{Umbreit:2009nt}
\bibinfo{author}{\bibfnamefont{S.}~\bibnamefont{Umbreit}},
  \bibinfo{author}{\bibfnamefont{J.~M.} \bibnamefont{Fregeau}},
  \bibnamefont{and} \bibinfo{author}{\bibfnamefont{F.~A.} \bibnamefont{Rasio}},
  \bibinfo{journal}{Astrophys. J.} \textbf{\bibinfo{volume}{719}},
  \bibinfo{pages}{915} (\bibinfo{year}{2010}).

\bibitem[{\citenamefont{Abadie et~al.}(2010{\natexlab{b}})}]{ratesdoc}
\bibinfo{author}{\bibfnamefont{J.}~\bibnamefont{Abadie}} \bibnamefont{et~al.}
  (\bibinfo{collaboration}{{LIGO Scientific Collaboration} and {Virgo
  Collaboration}}), \bibinfo{journal}{Class. Quant. Grav.}
  \textbf{\bibinfo{volume}{27}}, \bibinfo{pages}{173001}
  (\bibinfo{year}{2010}{\natexlab{b}}).

\bibitem[{\citenamefont{{Mandel} et~al.}(2008)\citenamefont{{Mandel}, {Brown},
  {Gair}, and {Miller}}}]{Mandel:2007rates}
\bibinfo{author}{\bibfnamefont{I.}~\bibnamefont{{Mandel}}},
  \bibinfo{author}{\bibfnamefont{D.~A.} \bibnamefont{{Brown}}},
  \bibinfo{author}{\bibfnamefont{J.~R.} \bibnamefont{{Gair}}},
  \bibnamefont{and} \bibinfo{author}{\bibfnamefont{M.~C.}
  \bibnamefont{{Miller}}}, \bibinfo{journal}{Astrophys. J.}
  \textbf{\bibinfo{volume}{681}}, \bibinfo{pages}{1431} (\bibinfo{year}{2008}).

\bibitem[{\citenamefont{Blackburn et~al.}(2008)}]{Blackburn:2008rt}
\bibinfo{author}{\bibfnamefont{L.}~\bibnamefont{Blackburn}}
  \bibnamefont{et~al.}, \bibinfo{journal}{Class. Quant. Grav.}
  \textbf{\bibinfo{volume}{25}}, \bibinfo{pages}{184004}
  (\bibinfo{year}{2008}).

\bibitem[{\citenamefont{Abadie et~al.}(in preparation)}]{VirgoDetChar}
\bibinfo{author}{\bibfnamefont{J.}~\bibnamefont{Abadie}} \bibnamefont{et~al.}
  (\bibinfo{collaboration}{LIGO Scientific Collaboration and Virgo
  Collaboration}), \bibinfo{journal}{Class. Quant. Grav.}  (\bibinfo{year}{in
  preparation}).

\bibitem[{\citenamefont{Klimenko et~al.}(2008)\citenamefont{Klimenko, Yakushin,
  Mercer, and Mitselmakher}}]{Klimenko:2007hd}
\bibinfo{author}{\bibfnamefont{S.}~\bibnamefont{Klimenko}},
  \bibinfo{author}{\bibfnamefont{I.}~\bibnamefont{Yakushin}},
  \bibinfo{author}{\bibfnamefont{A.}~\bibnamefont{Mercer}}, \bibnamefont{and}
  \bibinfo{author}{\bibfnamefont{G.}~\bibnamefont{Mitselmakher}},
  \bibinfo{journal}{Class. Quant. Grav.} \textbf{\bibinfo{volume}{25}},
  \bibinfo{pages}{114029} (\bibinfo{year}{2008}).

\bibitem[{\citenamefont{Klimenko et~al.}(2005)\citenamefont{Klimenko, Mohanty,
  Rakhmanov, and Mitselmakher}}]{Klimenko:2005wa}
\bibinfo{author}{\bibfnamefont{S.}~\bibnamefont{Klimenko}},
  \bibinfo{author}{\bibfnamefont{S.}~\bibnamefont{Mohanty}},
  \bibinfo{author}{\bibfnamefont{M.}~\bibnamefont{Rakhmanov}},
  \bibnamefont{and}
  \bibinfo{author}{\bibfnamefont{G.}~\bibnamefont{Mitselmakher}},
  \bibinfo{journal}{Phys. Rev. D} \textbf{\bibinfo{volume}{72}},
  \bibinfo{pages}{122002} (\bibinfo{year}{2005}).

\bibitem[{\citenamefont{Apostolatos et~al.}(1994)\citenamefont{Apostolatos,
  Cutler, Sussman, and Thorne}}]{Apostolatos:1994}
\bibinfo{author}{\bibfnamefont{T.~A.} \bibnamefont{Apostolatos}},
  \bibinfo{author}{\bibfnamefont{C.}~\bibnamefont{Cutler}},
  \bibinfo{author}{\bibfnamefont{G.~J.} \bibnamefont{Sussman}},
  \bibnamefont{and} \bibinfo{author}{\bibfnamefont{K.~S.}
  \bibnamefont{Thorne}}, \bibinfo{journal}{Phys.~Rev.~D}
  \textbf{\bibinfo{volume}{49}}, \bibinfo{pages}{6274} (\bibinfo{year}{1994}).

\bibitem[{\citenamefont{Blanchet et~al.}(1996)\citenamefont{Blanchet, Iyer,
  Will, and Wiseman}}]{Blanchet:1996pi}
\bibinfo{author}{\bibfnamefont{L.}~\bibnamefont{Blanchet}},
  \bibinfo{author}{\bibfnamefont{B.~R.} \bibnamefont{Iyer}},
  \bibinfo{author}{\bibfnamefont{C.~M.} \bibnamefont{Will}}, \bibnamefont{and}
  \bibinfo{author}{\bibfnamefont{A.~G.} \bibnamefont{Wiseman}},
  \bibinfo{journal}{Class. Quant. Grav.} \textbf{\bibinfo{volume}{13}},
  \bibinfo{pages}{575} (\bibinfo{year}{1996}).

\bibitem[{\citenamefont{Abadie et~al.}(2010{\natexlab{c}})}]{LSC:2010fk}
\bibinfo{author}{\bibfnamefont{J.}~\bibnamefont{Abadie}} \bibnamefont{et~al.},
  \bibinfo{journal}{Astrophy. J.} \textbf{\bibinfo{volume}{715}},
  \bibinfo{pages}{1453} (\bibinfo{year}{2010}{\natexlab{c}}).

\bibitem[{\citenamefont{Goggin}(2009)}]{Goggin:2009fv}
\bibinfo{author}{\bibfnamefont{L.~M.} \bibnamefont{Goggin}}, Ph.D. thesis,
  \bibinfo{school}{California Institute of Technology} (\bibinfo{year}{2009}).

\bibitem[{\citenamefont{Buonanno
  et~al.}(2009{\natexlab{b}})\citenamefont{Buonanno, Pan, Pfeiffer, Scheel,
  Buchman, and Kidder}}]{Buonanno:2009qa}
\bibinfo{author}{\bibfnamefont{A.}~\bibnamefont{Buonanno}},
  \bibinfo{author}{\bibfnamefont{Y.}~\bibnamefont{Pan}},
  \bibinfo{author}{\bibfnamefont{H.~P.} \bibnamefont{Pfeiffer}},
  \bibinfo{author}{\bibfnamefont{M.~A.} \bibnamefont{Scheel}},
  \bibinfo{author}{\bibfnamefont{L.~T.} \bibnamefont{Buchman}},
  \bibnamefont{and} \bibinfo{author}{\bibfnamefont{L.~E.}
  \bibnamefont{Kidder}}, \bibinfo{journal}{Phys. Rev. D}
  \textbf{\bibinfo{volume}{79}}, \bibinfo{pages}{124028}
  (\bibinfo{year}{2009}{\natexlab{b}}).

\bibitem[{\citenamefont{Ajith et~al.}(2007)}]{Ajith:2007qp}
\bibinfo{author}{\bibfnamefont{P.}~\bibnamefont{Ajith}} \bibnamefont{et~al.},
  \bibinfo{journal}{Class. Quant. Grav.} \textbf{\bibinfo{volume}{24}},
  \bibinfo{pages}{S689} (\bibinfo{year}{2007}).

\bibitem[{\citenamefont{Ajith et~al.}(2011)\citenamefont{Ajith, Hannam, Husa,
  Chen, Bruegmann, Dorband, Mueller, Ohme, Pollney, Reisswig
  et~al.}}]{Ajith:2009gj}
\bibinfo{author}{\bibfnamefont{P.}~\bibnamefont{Ajith}},
  \bibinfo{author}{\bibfnamefont{M.}~\bibnamefont{Hannam}},
  \bibinfo{author}{\bibfnamefont{S.}~\bibnamefont{Husa}},
  \bibinfo{author}{\bibfnamefont{Y.}~\bibnamefont{Chen}},
  \bibinfo{author}{\bibfnamefont{B.}~\bibnamefont{Bruegmann}},
  \bibinfo{author}{\bibfnamefont{N.}~\bibnamefont{Dorband}},
  \bibinfo{author}{\bibfnamefont{D.}~\bibnamefont{Mueller}},
  \bibinfo{author}{\bibfnamefont{F.}~\bibnamefont{Ohme}},
  \bibinfo{author}{\bibfnamefont{D.}~\bibnamefont{Pollney}},
  \bibinfo{author}{\bibfnamefont{C.}~\bibnamefont{Reisswig}},
  \bibnamefont{et~al.}, \bibinfo{journal}{Phys. Rev. Lett.}
  \textbf{\bibinfo{volume}{106}}, \bibinfo{pages}{241101}
  (\bibinfo{year}{2011}).

\bibitem[{\citenamefont{Pankow}(2011)}]{Pankow:2011aa}
\bibinfo{author}{\bibfnamefont{C.}~\bibnamefont{Pankow}}, Ph.D. thesis,
  \bibinfo{school}{University of Florida} (\bibinfo{year}{2011}),
  \urlprefix\url{http://uf.catalog.fcla.edu/uf.jsp?st=UF005295299&ix=pm&I=0&V=D&pm=1}.

\bibitem[{\citenamefont{Pankow and Klimenko}(2011)}]{Klimenko:2011ab}
\bibinfo{author}{\bibfnamefont{C.}~\bibnamefont{Pankow}} \bibnamefont{and}
  \bibinfo{author}{\bibfnamefont{S.}~\bibnamefont{Klimenko}}
  (\bibinfo{publisher}{Presented at the Gravitational Wave Physics and
  Astronomy Workshop}, \bibinfo{year}{2011}).

\bibitem[{\citenamefont{Klimenko and C}(2012)}]{Klimenko:2012aa}
\bibinfo{author}{\bibfnamefont{S.}~\bibnamefont{Klimenko}} \bibnamefont{and}
  \bibinfo{author}{\bibfnamefont{P.}~\bibnamefont{C}}, \bibinfo{type}{T1200169}
  \bibinfo{number}{v1}, \bibinfo{institution}{LIGO Project}
  (\bibinfo{year}{2012}),
  \urlprefix\url{https://dcc.ligo.org/cgi-bin/DocDB/ShowDocument?docid=90004}.

\bibitem[{\citenamefont{Abadie
  et~al.}(2010{\natexlab{d}})}]{Collaboration:2010fl}
\bibinfo{author}{\bibfnamefont{J.}~\bibnamefont{Abadie}} \bibnamefont{et~al.},
  \bibinfo{journal}{Nucl. Instrum. Meth.} \textbf{\bibinfo{volume}{624}},
  \bibinfo{pages}{223} (\bibinfo{year}{2010}{\natexlab{d}}).

\bibitem[{\citenamefont{Accadia et~al.}(2010)}]{collaboration:2010up}
\bibinfo{author}{\bibfnamefont{T.}~\bibnamefont{Accadia}} \bibnamefont{et~al.}
  (\bibinfo{collaboration}{Virgo Collaboration}), \bibinfo{journal}{J. Phys.
  Conf. Ser.} \textbf{\bibinfo{volume}{228}}, \bibinfo{pages}{012015}
  (\bibinfo{year}{2010}).

\bibitem[{\citenamefont{Pan et~al.}(2011)\citenamefont{Pan, Buonanno, Boyle,
  Buchman, Kidder, Pfeiffer, and Scheel}}]{Pan:2011fk}
\bibinfo{author}{\bibfnamefont{Y.}~\bibnamefont{Pan}},
  \bibinfo{author}{\bibfnamefont{A.}~\bibnamefont{Buonanno}},
  \bibinfo{author}{\bibfnamefont{M.}~\bibnamefont{Boyle}},
  \bibinfo{author}{\bibfnamefont{L.~T.} \bibnamefont{Buchman}},
  \bibinfo{author}{\bibfnamefont{L.~E.} \bibnamefont{Kidder}},
  \bibinfo{author}{\bibfnamefont{H.~P.} \bibnamefont{Pfeiffer}},
  \bibnamefont{and} \bibinfo{author}{\bibfnamefont{M.~A.} \bibnamefont{Scheel}}
  (\bibinfo{year}{2011}), \eprint{1106.1021v1},
  \urlprefix\url{http://arxiv.org/abs/1106.1021v1}.

\bibitem[{\citenamefont{{O'Shaughnessy}
  et~al.}(2010)\citenamefont{{O'Shaughnessy}, {Vaishnav}, {Healy}, and
  {Shoemaker}}}]{2010PhRvD..82j4006O}
\bibinfo{author}{\bibfnamefont{R.}~\bibnamefont{{O'Shaughnessy}}},
  \bibinfo{author}{\bibfnamefont{B.}~\bibnamefont{{Vaishnav}}},
  \bibinfo{author}{\bibfnamefont{J.}~\bibnamefont{{Healy}}}, \bibnamefont{and}
  \bibinfo{author}{\bibfnamefont{D.}~\bibnamefont{{Shoemaker}}},
  \bibinfo{journal}{\prd} \textbf{\bibinfo{volume}{82}}, \bibinfo{eid}{104006}
  (\bibinfo{year}{2010}), \eprint{1007.4213}.

\bibitem[{\citenamefont{Biswas et~al.}(2009)\citenamefont{Biswas, Brady,
  Creighton, and Fairhurst}}]{Biswas:2007vn}
\bibinfo{author}{\bibfnamefont{R.}~\bibnamefont{Biswas}},
  \bibinfo{author}{\bibfnamefont{P.~R.} \bibnamefont{Brady}},
  \bibinfo{author}{\bibfnamefont{J.~D.~E.} \bibnamefont{Creighton}},
  \bibnamefont{and}
  \bibinfo{author}{\bibfnamefont{S.}~\bibnamefont{Fairhurst}},
  \bibinfo{journal}{Class. Quant. Grav.} \textbf{\bibinfo{volume}{26}},
  \bibinfo{pages}{175009} (\bibinfo{year}{2009}).

\bibitem[{\citenamefont{Reisswig et~al.}(2009)\citenamefont{Reisswig, Husa,
  Rezzolla, Dorband, Pollney, and Seiler}}]{Reisswig:2009fk}
\bibinfo{author}{\bibfnamefont{C.}~\bibnamefont{Reisswig}},
  \bibinfo{author}{\bibfnamefont{S.}~\bibnamefont{Husa}},
  \bibinfo{author}{\bibfnamefont{L.}~\bibnamefont{Rezzolla}},
  \bibinfo{author}{\bibfnamefont{E.~N.} \bibnamefont{Dorband}},
  \bibinfo{author}{\bibfnamefont{D.}~\bibnamefont{Pollney}}, \bibnamefont{and}
  \bibinfo{author}{\bibfnamefont{J.}~\bibnamefont{Seiler}},
  \bibinfo{journal}{Phys. Rev. D} \textbf{\bibinfo{volume}{80}},
  \bibinfo{pages}{124026} (\bibinfo{year}{2009}),
  \urlprefix\url{http://link.aps.org/doi/10.1103/PhysRevD.80.124026}.

\bibitem[{\citenamefont{Vidakovic}(1999)}]{Vidakovic:1999fk}
\bibinfo{author}{\bibfnamefont{B.}~\bibnamefont{Vidakovic}},
  \emph{\bibinfo{title}{Statistical Modeling by Wavelets}}
  (\bibinfo{publisher}{Wiley}, \bibinfo{year}{1999}).

\end{thebibliography}
